\documentclass[12pt]{article}
\usepackage{latexsym}
\usepackage{amsmath}
\usepackage{amssymb}
\catcode`\@=11
\newcount\hour
\newcount\minute
\newtoks\amorpm \hour=\time\divide\hour by 60\minute
=\time{\multiply\hour by 60 \global\advance\minute by-\hour}
\edef\standardtime{{\ifnum\hour<12 \global\amorpm={am}%
        \else\global\amorpm={pm}\advance\hour by-12 \fi
        \ifnum\hour=0 \hour=12 \fi
        \number\hour:\ifnum\minute<10
        0\fi\number\minute\the\amorpm}}
\edef\militarytime{\number\hour:\ifnum\minute<10
0\fi\number\minute}
\def\draftlabel#1{{\@bsphack\if@filesw {\let\thepage\relax
   \xdef\@gtempa{\write\@auxout{\string
      \newlabel{#1}{{\@currentlabel}{\thepage}}}}}\@gtempa
   \if@nobreak \ifvmode\nobreak\fi\fi\fi\@esphack}
        \gdef\@eqnlabel{#1}}
\def\@eqnlabel{}
\def\@vacuum{}
\def\marginnote#1{}
\def\draftmarginnote#1{\marginpar{\raggedright\scriptsize\tt#1}}
\def\draft{
        \pagestyle{plain}
        \overfullrule=2pt
        \oddsidemargin -.1truein
        \def\@oddhead{\sl \phantom{\today\quad\militarytime} \hfil
        \smash{\Large\sl DRAFT} \hfil \today\quad\militarytime}
        \let\@evenhead\@oddhead
        \let\label=\draftlabel
        \let\marginnote=\draftmarginnote
        \def\ps@empty{\let\@mkboth\@gobbletwo
        \def\@oddfoot{\hfil \smash{\Large\sl DRAFT} \hfil}
        \let\@evenfoot\@oddhead}
        \def\@eqnnum{(\theequation)\rlap{\kern\marginparsep\tt\@eqnlabel}%
        \global\let\@eqnlabel\@vacuum}  }

\renewcommand{\theequation}{\thesection.\arabic{equation}}
\renewcommand{\thefootnote}{\fnsymbol{footnote}}
\newcommand{\newsection}{    
\setcounter{equation}{0}\section}
\def\appendix#1{\addtocounter{section}{1}\setcounter{equation}{0}
\renewcommand{\thesection}{\Alph{section}}
\section*{Appendix \thesection\protect\indent \parbox[t]{11.15cm}{#1}}
\addcontentsline{toc}{section}{Appendix \thesection\ \ \ #1}}

\def \bi{\bibitem}
\def \la {\label}

\def \b {\beta}

\def \t {\theta}
\def \s{\sigma}
\def \d {\partial}

\def\be{\begin{equation}}
\def\ee{\end{equation}}

\def\nat {{\natural}}

\newcommand{\cont}[1]{{}_{#1}{}^{#1}}

\catcode`\@=11

\def\bea{\begin{eqnarray}}
\def\eea{\end{eqnarray}}
\def\beann{\begin{eqnarray*}}
\def\eeann{\end{eqnarray*}}
\def\beq{\begin{equation}}
\def\eeq{\end{equation}}
\def\ba{\begin{array}}
\def\ea{\end{array}}
\def\ben{\begin{enumerate}}
\def\een{\end{enumerate}}
 \def \l {\lambda}
 \def\m {\mu}
\def\s {\sigma }
 \def \la {\label}
 \def\be{\begin{equation}}
\def\ee{\end{equation}}

\def \la {\label}

\def \r {\rho}

\font\mybb=msbm10 at 11pt

\def\bb#1{\hbox{\mybb#1}}

\def\bR {\bb{R}}

\def\e  {\epsilon}

\def \t {\theta}

\def \ee {\epsilon}

\def \g {\gamma}
\def \bi{\bibitem}
\def\a{\alpha }

\def \s {\sigma}

\def \r {\rho}
\def \d {\delta}

\def \l {\lambda}
\def \m {\mu}
\def \g {\gamma}
\def \n {\nu}

\def \b {\beta}

\def \tone {(T^1)}
\def \ttwo {(T^2)}
\def \tthree {(T^3)}
\def \tfour {(T^4)}
\def \tfive {(T^5)}
\def \bn {{\bar{\beta}}}

\def\be{\begin{equation}}
\def\ee{\end{equation}}

\def \bi {\bibitem}
\def \la{\label}

\def \t {\tau}

\begin{document}
\date{November 2002}
\begin{titlepage}
\begin{center}
\hfill hep-th/0610331 \\
\hfill KUL-TF-06/27 \\
\hfill UB-ECM-PF-06-31 \\

\vspace{3.0cm} {\Large \bf $N=31$, $D=11$} \\[.2cm]

\vspace{1.5cm}
 {\large  U. Gran$^1$, J. Gutowski$^2$,  G. Papadopoulos$^3$ and D. Roest$^4$}

\vspace{0.5cm}

${}^1$ Institute for Theoretical Physics, K.U. Leuven\\
Celestijnenlaan 200D\\
B-3001 Leuven, Belgium\\

\vspace{0.5cm}
${}^2$ DAMTP, Centre for Mathematical Sciences\\
University of Cambridge\\
Wilberforce Road, Cambridge, CB3 0WA, UK

\vspace{0.5cm}
${}^3$ Department of Mathematics\\
King's College London\\
Strand\\
London WC2R 2LS, UK\\

\vspace{0.5cm}
${}^4$ Departament Estructura i Constituents de la Materia \\
    Facultat de F\'{i}sica, Universitat de Barcelona \\
    Diagonal, 647, 08028 Barcelona, Spain \\

\end{center}

\vskip 1.5 cm
\begin{abstract}

We show that eleven-dimensional supergravity backgrounds with thirty one supersymmetries, $N=31$, admit an additional Killing spinor
and so they are locally
isometric to maximally supersymmetric ones. This rules out the existence of simply connected eleven-dimensional
supergravity preons. We also show that \linebreak $N=15$  solutions of type I supergravities are locally isometric to Minkowski spacetime.

\end{abstract}
\vskip 0.8cm

\line(1,0){50}

{\small {The laconic title has been inspired by that of \cite{howe}.}}
\end{titlepage}
\newpage
\setcounter{page}{1}
\renewcommand{\thefootnote}{\arabic{footnote}}
\setcounter{footnote}{0}

\setcounter{section}{0}
\setcounter{subsection}{0}
\newsection{Introduction}

The spinorial geometry technique is an effective tool to solve the Killing spinor equations
of supergravity theories \cite{ggp}. It is based on the use of gauge symmetry of the Killing spinor equations,
on a description of spinors in terms of forms and on an oscillator basis in the space of spinors.
Recently, it has been  adapted to investigate backgrounds with near maximal number of
supersymmetries. In particular it was found that IIB supergravity backgrounds with
31 supersymmetries, $N=31$, are maximally supersymmetric \cite{iibpreons}.
This was extended in \cite{bandos}, using a different method, to show that
IIA $N=31$ supergravity backgrounds are also maximally supersymmetric. Later
the spinorial geometry approach was applied to lower-dimensional
supergravities\footnote{Technical innovations developed for this paper have been applied to establish the results
in lower dimensions. This project precedes those in \cite{jgutowski}.}  with similar results \cite{jgutowski}.

In this paper, we shall show that the $N=31$ backgrounds of eleven-dimensional supergravity \cite{cjs}, termed as preons in \cite{bandospreons},
admit locally an additional Killing spinor and so they are
 maximally supersymmetric.
 Although this result is similar to those in type II supergravities mentioned above,
 there are some differences. To establish the type II results, the algebraic Killing spinor equations of type II supergravities
 have been instrumental. The remaining parallel transport equations were not
 explicitly solved and instead the result
 followed by an indirect argument. The eleven-dimensional supergravity does not have an algebraic Killing spinor equation.
 So to show that the $N=31$ backgrounds are locally isometric to the maximally supersymmetric ones, the parallel transport
 equation
 \bea
 {\cal D}_A\epsilon^r=0~,~~~r=1,\dots,31~,
 \la{kseqn}
 \eea
  has to be solved explicitly. For this, one investigates the first integrability condition
  \bea
  {\cal R}_{AB} \e^r=[{\cal D}_A, {\cal D}_B]\e^r=0~,
  \la{supcurveqn}
  \eea
 where ${\cal R}$ is the supercovariant curvature. The stability subgroup, ${\rm Stab}(\e)$, of $31$ spinors $\e$ in the
 holonomy group $SL(32,\bR)$ is ${\rm Stab}(\e)=\bR^{31}$ \cite{hull, duff, gpdta}. Thus the integrability condition leaves
 undetermined 31
 components of ${\cal R}$ represented by 31 two-forms $u^r_{AB}$. The task is to show that these components vanish as well
 and so the (reduced) holonomy of the supercovariant connection for $31$ Killing spinors is in fact $\{1\}$. To do this, we shall use
 a modification of the procedure outlined in \cite{iibpreons} which utilizes the normal $\nu$ of the Killing spinors $\e^r$
 and which is explained in the next section. Then we shall use
 the Bianchi identities, the field equations and the explicit expression of ${\cal R}$ in terms of the fields
 of eleven-dimensional supergravity to show  that  the supercovariant curvature vanishes, ${\cal R}=0$. The latter condition
 is sufficient to demonstrate that the $N=31$ backgrounds are locally isometric
 to the maximally supersymmetric ones. The maximally supersymmetric backgrounds  have been
 classified in \cite{jfgpa}, and has been shown to be locally isometric
 to Minkowski space $\bR^{10,1}$, the Freund-Rubin \cite{frubin}
 spaces $AdS_4\times S^7$ and $AdS_7\times S^4$, and the Kowalski-Glikman plane wave \cite{kglikman}, see also \cite{jfofgpc}.

On non-simply connected spacetimes, the vanishing of the supercovariant curvature, ${\cal R}=0$, does not always imply the existence
of 32 linearly independent solutions for the parallel transport equation (\ref{kseqn}).
There is also the additional subtlety
of the existence of different spin structures on non-simply connected spacetimes.
Since we  show
that the $N=31$ backgrounds are locally isometric to the maximally supersymmetric ones, one may be able to construct
 $N=31$ supersymmetric backgrounds by identifying one of the maximal supersymmetric ones
with a discrete subgroup of its symmetry group. A large class of such backgrounds were constructed in  \cite{jfof} after identification with
 a cyclic subgroup of the symmetry group.
 None of these preserve 31 supersymmetries\footnote{We thank J.~Figueroa O'Farrill for helpful discussions
on this point.}. This may indicate that  non-simply connected backgrounds with $N=31$ supersymmetries do not exist
 but some further investigation is required to establish this.
 The absence of $N=31$ supersymmetric backgrounds will be  in agreement with a conjecture in \cite{duffviii} which
 was formulated under the assumption that the Killing spinors must lie in certain representations of subgroups
 of $Spin(10,1)$.

We also show that the $N=15$ solutions in type I supergravities are locally maximally supersymmetric. This easily follows
from our result in IIB \cite{iibpreons} and the investigation of the Killing spinor equations
of the heterotic supergravity in \cite{phgpug}.

The paper has been organized as follows. In section two, we explain the procedure we use to investigate backgrounds
with 31 supersymmetries and collect some useful formulae. In addition, we show that there are two cases to
examine depending on the stability subgroup of the normal $\nu$ to the 31 Killing  spinors in $Spin(10,1)$. In section three,
we investigate the $N=31$ backgrounds whose normal $\nu$ has stability subgroup $SU(5)$,
and in section four we examine the $N=31$ backgrounds whose normal $\nu$ has stability subgroup
$(Spin(7)\ltimes\bR^8)\times\bR$. In both cases, we establish that the $N=31$ backgrounds are locally isometric
to the maximally supersymmetric ones. In section five, we examine the $N=15$ backgrounds of type I supergravities.
 In section six, we present our conclusions.

\newsection{Supercurvature and Killing spinors}

As we have mentioned, a consequence of the Killing spinor equations is the integrability condition
(\ref{supcurveqn}). In \cite{iibpreons}, it was proposed to solve this condition directly. This has been
facilitated by first using the gauge symmetry of the Killing spinor equations to choose the direction
of the normal spinor $\nu$ of the $N=31$ Killing spinors. In turn the gauge symmetry orients the hyperplane
of the 31 Killing spinors along particular directions. This simplifies the expression for the Killing spinors
and then using spinorial geometry the condition ${\cal R}\e^r=0$  gives rise to
 a linear system for the various components of the supercurvature. The linear system can be solved to
 give the conditions on ${\cal R}$ imposed by supersymmetry. Although this is the original way that
 we have tackled the problem,  it turns out there is a simpler way to explore the integrability condition (\ref{supcurveqn}). For this
let $\e^r$, $r=1, \dots, N$, be a basis in the space of Killing spinors and extend it as $(\e^r, \tilde \e^q)$, $q=N+1,\dots, 32$
 to a basis in the space of spinors.  Then observe that the supercovariant curvature
 for a background with $N$ Killing spinors can be written as
 \bea
{\cal R}_{MN,ab}= U_{MN,rp}\,\e^r_a
\nu^p_b+U_{MN,pq}\,\tilde\e^p_a \nu^q_b~,
\la{supcurv}
\eea
where $\nu^p$ are normal to the Killing spinors, $a,b=1,\dots,32$ are spinor indices,
\bea
B(\e^r, \nu^q)=0~,
\la{normcon}
\eea
and $U$ are spacetime dependent two-forms. (Throughout this paper we use the conventions of \cite{uggpdr}.) Clearly (\ref{supcurv}) satisfies
the integrability condition (\ref{supcurveqn}) because of (\ref{normcon}).
Since the holonomy of the supercovariant connection is contained in
$SL(32,\bR)$, one finds that
\bea
U_{MN,pq}\, B(\tilde\e^p,\nu^q)=0~. \la{supcon12}
\eea
Taking into account this condition, the number of independent
two-forms $U$ that appear in (\ref{supcurv}) is $32^2-32N-1$ which
is the dimension of the stability subgroup $SL(32-N,\bR)\ltimes
(\oplus_N \bR^{32-N})$ of $N$ spinors in $SL(32,\bR)$, see \cite{hull, duff, gpdta, gpdtb}.

In many cases of interest, the Killing spinors can be (locally) expressed in terms of a convenient basis $\eta^r$ as
\bea
\e^r= f^r{}_s \eta^s~,
\eea
where $f$ is an $N\times N$ invertible matrix  of  spacetime functions. If $(\eta^r, \tilde\eta^p)$ is a basis
in the space of spinors, then (\ref{supcurv}) can be written as
\bea
{\cal R}_{MN,ab}= u_{MN,rp}\,\eta^r_a
\nu^p_b+u_{MN,pq}\,\tilde\eta^p_a \nu^q_b~,
\la{supcurv12}
\eea
where $U$ and $u$ are related in a straightforward way.

The supercurvature can be written as
\bea
{\cal R}_{MN,ab}=\sum^5_{k=1} {1\over k!}(T_{MN}^k)_{A_1A_2\dots A_k}
(\Gamma^{A_1A_2\dots A_k})_{ab}~,
\eea
where $T^k$ depends on the frame $e$ and four-form field strength $F$ of eleven-dimensional supergravity. The relevant
expressions\footnote{There are some apparent typos in the expression for ${\cal R}$ in \cite{jfgpa}.} can be found in \cite{biran, jfgpa}.
It is also known that
\bea
\eta_a \theta_b=\frac{1}{32}\sum^5_{k=0} {(-1)^{k+1}\over k!} B(\eta,
\Gamma_{A_1A_2\dots A_k}\theta)\,  (\Gamma^{A_1A_2\dots
A_k})_{ab}~.
\eea
This in particular  implies that
\bea
(T_{MN}^k)_{A_1A_2\dots A_k}=\frac{(-1)^{k+1}}{32} [u_{MN,ip}\,
B(\eta^i, \Gamma_{A_1A_2\dots A_k}\nu^p)+u_{MN,pq}\, B(\tilde\eta^p,
\Gamma_{A_1A_2\dots A_k}\nu^q)] \la{supcurv2}
\eea
subject to the condition (\ref{supcon12}) which can now be rewritten as
\bea
u_{MN,pq}\, B(\tilde\eta^p,\nu^q)=0~. \la{supcon2}
\eea
The conditions (\ref{supcurv2})  are equivalent to those that arise  from the direct solution of the
integrability condition (\ref{supcurveqn}). The advantage is that  (\ref{supcurv2}) is more easy to  handle.

 The conditions (\ref{supcurv2}) and (\ref{supcon2}) can be easily adapted to backgrounds
 with 31 supersymmetries to find
\bea
(T_{MN}^k)_{A_1A_2\dots A_k}=\frac{(-1)^{k+1}}{32} u_{MN,r}\, B(\eta^r,
\Gamma_{A_1A_2\dots A_k}\nu)~.
\label{T-u-relation}
\eea
The second term in the r.h.s of (\ref{supcurv2}) vanishes because of
(\ref{supcon2}). This formula is consistent with the
requirement that the holonomy of the supercovariant connection for
$N=31$ configurations is in $\bR^{31}$.

Apart from the restrictions required by holonomy and described above, the supercovariant curvature ${\cal R}$ satisfies additional
conditions which arise from the field equations, the Bianchi identities of the Riemann curvature $R$ of the spacetime and of
the four-form field strength $F$ of eleven-dimensional supergravity, and the explicit expression of the components of ${\cal R}$
in terms of the fields. We can derive some of them
by observing that  $\Gamma^N{\cal R}_{MN}$ is a  linear combination of field equations and Bianchi identities, and so
it necessarily vanishes. In turn this leads to the vanishing of the following linear combinations
of the components of ${\cal R}$:
  \bea
    && (T^1_{MN})^N = 0~,~~~  (T^2_{MN})_P{}^N = 0~,~~~(T^1_{MP_1})_{P_2} + \tfrac{1}{2} (T^3_{MN})_{P_1 P_2}{}^N =  0~, \cr
    && (T^2_{M[P_1})_{P_2 P_3]} - \tfrac{1}{3} (T^4_{MN})_{P_1 P_2 P_3}{}^N =  0~,~~~
    (T^3_{M[P_1})_{P_2 P_3 P_4]} + \tfrac{1}{4} (T^5_{MN})_{P_1 \cdots P_4}{}^N =  0~, \cr
    && (T^4_{M[P_1})_{P_2 \cdots P_5]} - \frac{1}{5 \cdot 5!} \epsilon_{P_1 \cdots P_5}{}^{Q_1 \cdots Q_6}
      (T^5_{M Q_1})_{Q_2 \cdots Q_6}  =  0~.
 \label{field-eqs}
 \eea
The second and third  of these equations are consequences of the Einstein
and $F$ field equations, respectively.
We shall also use the additional conditions
 \bea
  && (T^1_{MN})_P = (T^1_{[MN})_{P]} \,, \qquad
  (T^2_{MN})_{PQ} = (T^2_{PQ})_{MN} \,, \qquad
  (T^3_{[MN})_{PQR]} = 0 \,,
 \label{T-constraints}
 \eea
which can be easily derived by inspecting the explicit expressions of $T^k$ in terms of the physical fields in \cite{jfgpa}
and by using
the Bianchi identity of $F$. Observe that the first condition in (\ref{field-eqs}) is a consequence
of the first condition in (\ref{T-constraints}). It will turn out that (\ref{field-eqs}), (\ref{T-constraints}), the expression of $T^k$ in terms of the
physical fields  and
the conditions (\ref{T-u-relation}) are sufficient for the proof that we shall present.

It has been known for some time that there are two kinds of orbits of $Spin(10,1)$ in the space of Majorana
spinors of eleven-dimensional supergravity. One has stability subgroup
$SU(5)$ and the other has stability subgroup $(Spin(7)\ltimes\bR^8)\times \bR$ \cite{bryant, figueroab}.
Therefore, there are two cases of $N=31$ backgrounds to
explore depending on in which orbit the normal $\nu$ of the Killing spinors lies. This is similar to the
$N=31$ IIB backgrounds in \cite{iibpreons}.  The Killing spinor equations for the associated $N=1$ eleven-dimensional backgrounds
have been solved in \cite{gjs}.
 We shall  investigate the
two $N=31$ cases separately.

\newsection{$SU(5)$-invariant normal}

\subsection{Integrability conditions}

To derive the conditions that the integrability of the Killing spinor equations imposes on the
supercurvature,
without loss of generality, we choose the normal of the 31 Killing spinors  as
\bea
\nu=1+e_{12345}~,
\eea
in the ``time-like'' spinor basis of  \cite{uggpdr}.
Then the Killing spinors can be written as
\bea
\epsilon^r=f^r{}_s \eta^s~,
\eea
where $f$ is a $31\times 31$ invertible matrix of real spacetime functions and $\eta^s$
is a basis of 31 linearly independent Majorana spinors. This basis  can be chosen as
\bea
&&\eta^0=\nu=1+e_{12345}~,
\cr
&&\eta^{k}=-i(e_k-{1\over 4!}
\epsilon_k{}^{q_1q_2q_3q_4} e_{q_1q_2q_3q_4})~,~~~
\eta^{5+k}=e_k+{1\over 4!} \epsilon_k{}^{q_1q_2q_3q_4}
e_{q_1q_2q_3q_4}~,
\cr
&&\eta^{kl}=e_{kl}-{1\over 3!} \epsilon_{kl}{}^{q_1q_2q_3}
e_{q_1q_2q_3}~,~~~\hat\eta^{kl}=i (e_{kl}+{1\over 3!}
\epsilon_{kl}{}^{q_1q_2q_3} e_{q_1q_2q_3})~,
 \label{Maj-basis}
\eea
where $k,l=1,\ldots,5$.  Observe that the linearly independent Majorana spinor $i-ie_{12345}$ is not orthogonal to the normal $\nu$
and so it has been excluded from the basis. It is
convenient for what follows to set  $\theta^0=\eta^0$
and then choose a `holomorphic' basis for the rest of the
spinors as
\bea
&&\theta^\a=\eta^\a+i
\eta^{\a+5}~,~~~\theta^{\bar\a}=\eta^{\a}-i \eta^{\a+5}~,
\cr
&&\theta^{\a\b}=\eta^{\a\b}+i
\hat\eta^{\a\b}~,~~~\theta^{\bar\a\bar\b}=\eta^{\a\b}-i
\hat\eta^{\a\b}~,
 \label{SU(5)-basis}
\eea
i.e.~decompose ${\bf 31}={\bf 1}\oplus {\bf 5}\oplus \bar {\bf 5}\oplus {\bf 10}\oplus \bar {\bf 10}$ in
$SU(5)$ representations and so $r=(0,\a,\bar\a,\a\b,\bar\a\bar\b)$. This has the advantage that the conditions (\ref{T-u-relation}) can be expressed
in an $SU(5)$ covariant manner.

To find the conditions that arise from the integrability condition
(\ref{T-u-relation}), it is necessary to compute the spinor bi-linear forms. These have been presented in
appendix A. It is then straightforward to see that (\ref{T-u-relation}) implies that
 \bea
  u_{MN,0} = - (T^1_{MN})_0 \,, \quad u_{MN,_\a} = (T^1_{MN})_\a \,, \quad u_{MN, \a \b} = \frac{1}{2
    \sqrt{2}} i (T^2_{MN})_{\a \b}~.
 \eea
In addition, (\ref{T-u-relation}) gives
\bea
\label{block1}
\ttwo_{0\a} &=& i \tone_\a \,,
\cr
\ttwo_{\a \bar{\b}} &=& - i g_{\a \bar{\b}} \tone_0 \,,
\cr
\tthree_{0 \b_1 \b_2} &=& - i \ttwo_{\b_1 \b_2} \,,
\cr
\tthree_{0 \a \bar{\b}} &=&0 \,,
\cr
\tthree_{\bn_1 \bn_2 \bn_3} &=& \tfrac{1}{2} \sqrt{2} \epsilon_{\bn_1 \bn_2
  \bn_3}{}^{\a_1 \a_2} \ttwo_{\a_1 \a_2} \,,
\cr
\tthree_{\a \bn_1 \bn_2} &=& -2 \tone_{[\bn_1} g_{\bn_2] \a} \,,
\cr
\tfour_{0 \bn_1 \bn_2 \bn_3} &=&  - \tfrac{1}{2} \sqrt{2} i \epsilon_{\bn_1 \bn_2
  \bn_3}{}^{\a_1 \a_2} \ttwo_{\a_1 \a_2} \,,
\cr
\tfour_{0\a \bn_1 \bn_2} &=& 2 i \tone_{[\bn_1} g_{\bn_2] \a} \,,
\cr
\tfour_{\a_1 \a_2 \a_3 \a_4} &=& - 2 \sqrt{2} \epsilon_{\a_1 \a_2 \a_3
  \a_4}{}^\bn \tone_\bn  \,,
\cr
\tfour_{\a \bn_1 \bn_2 \bn_3} &=& 3 \ttwo_{[\bn_1 \bn_2} g_{\bn_3] \a} \,,
\cr
\tfour_{\a_1 \a_2 \bar\g_1 \bar\g_2} &=&0 \,,
\cr
\tfive_{0 \a_1 \a_2 \a_3 \a_4} &=& 2 \sqrt{2} i \epsilon_{\a_1 \a_2 \a_3
  \a_4}{}^\bn \tone_\bn \,,
\cr
\tfive_{0 \a \bn_1 \bn_2 \bn_3} &=& 3 i \ttwo_{[\bn_1 \bn_2} g_{\bn_3] \a} \,,
\cr
\tfive_{0\a_1 \a_2 \bn_1 \bn_2} &=& - 2   \tone_0 g_{\a_1 [\bn_1}
g_{\bn_2] \a_2} \,,
\cr
(T^5)_{\a_1 \cdots \a_5} &=& 2 \sqrt{2} i \epsilon_{\a_1 \cdots \a_5} \tone_0 \,,
\cr
\tfive_{\a \bn_1 \bn_2 \bn_3 \bn_4} &=& - \sqrt{2} \epsilon_{\bn_1 \bn_2 \bn_3
  \bn_4}{}^{\g} \ttwo_{\a \g} \,,
\cr
\tfive_{\a_1 \a_2 \bn_1 \bn_2 \bn_3} &=& -6 \tone_{[\bn_1} g_{|\a_1| \bn_2} g_{\bn_3] \a_2} \,,
\la{holcon}
\eea
where we have suppressed the two-form indices.
Observe that all $T^k$ have been expressed in terms of $T^1$ and $T^2$.
The above conditions are equivalent to the integrability condition ${\cal R}\e^r=0$. Clearly, they do not imply that
${\cal R}=0$. It now remains to impose the conditions  (\ref{field-eqs}) and (\ref{T-constraints}).

\subsection{Solving the conditions}

We shall first show using (\ref{field-eqs}),  (\ref{T-constraints}) and (\ref{holcon})
that $T^1$ vanishes. For this observe that (\ref{holcon}) together with the skew-symmetry of $T^1$, (\ref{T-constraints}), implies
that
 \bea
  && (T^2_{0 \a})_{\b \bar \g} = - i T^1_{0 \a 0 }\, g_{\b \bar \g} = 0 \,.
 \eea
Using the symmetry property of $T^2$ in (\ref{T-constraints}), this
 leads to
 \bea
  && (T^2_{\b \bar \g })_{0 \a} = i T^1_{\a \b \bar \g} = 0 \,,
 \eea
and hence the $(2,1)$ and $(1,2)$ parts of $T^1$ vanish. Turning to
the field equations, using the Einstein equation in (\ref{field-eqs}) and (\ref{holcon}), we find that
 \bea
  && T^1_{0 \a}{}^\a = T^1_{0 \a \b} = 0 \,, \qquad
    (T^2_{\a \g})_{\bar \b}{}^\g = -2 i T^1_{0 \a \bar \b} \,.
 \label{Einstein-eq}
 \eea
Similarly, the gauge field equation in (\ref{field-eqs}) leads to
 \bea
  && T^1_{\a \b \g} =0\, .
\label{gauge-eq}
 \eea
The only remaining component of $T^1$ is the traceless part of
$T^1_{0 \a \bar
  \b}$. Its relation to $T^2$ is
 \bea
  (T^2_{\a \bar \b})_{\g \bar \d} = - i T^1_{0 \a \bar
  \b}\, g_{\g \bar \d} \,.
 \eea
Tracing this expression with $g^{\g \bar \d}$ and using the symmetry
in the two pairs of indices of $T^2$, this gives
 \bea
  (T^2_{\a \bar \b})_{\g}{}^\g = - 5 i T^1_{0 \a \bar
  \b} = - i g_{\a \bar \b} T^1_{0 \g}{}^\g =0\,.
 \eea
The last equality follows from  (\ref{Einstein-eq}). Therefore $T^1=0$.

It remains to show $T^2=0$ as well. An inspection of the conditions we have derived above reveal that
the only non-vanishing components are $(T^2_{\a \b})_{\g
\d}$ and $(T^2_{\a \b})_{\bar \g \bar \d}$. The former vanishes because of the Bianchi identity of $T^3$, (\ref{T-constraints}),
involving skew-symmetry in two holomorphic and three anti-holomorphic indices,   and the the relation of $T^3$ to $T^2$ in (\ref{holcon}).
To continue, first observe that from (\ref{Einstein-eq}) and $T^1=0$, we find that
\bea
(T^2_{\a \g})_{\bar \b}{}^\g=0~.
\la{trcone}
\eea
Next we shall use the expression of $T^1$ and $T^3$ in terms of the fluxes $F$ which can be found in \cite{jfgpa}.
The condition $T^1=0$ implies that $F\wedge F=0$ which in turn implies that
\bea
   (T^3_{MN})_{PQR} = \tfrac{1}{6} (\nabla_M F_{NPQR} - \nabla_N F_{MPQR}) \,.
  \eea
 Now consider the case where all five indices are holomorphic. This
 component of $T^3$ is subject to two additional conditions. The first follows
 from the Bianchi identity for the gauge field, which states that
 \bea
  (T^3_{[\a \b_1})_{\b_2 \b_3 \b_4]} = \tfrac{1}{15} (\nabla_{\a} F_{\b_1
    \cdots \b_4} + 4 \nabla_{[\b_1}
  F_{\b_2 \b_3 \b_4] \a }) = 0 \,.
  \la{ttone}
 \eea
The second condition follows from the relation between $T^3$ and
$T^2$ in (\ref{block1}) and the trace condition on $T^2$ in
(\ref{trcone}). It implies that
 \bea
 (T^3_{\a [\b_1})_{\b_2 \b_3 \b_4]} = \tfrac{1}{6} (\nabla_{\a} F_{\b_1
    \cdots \b_4} + \nabla_{[\b_1}
  F_{\b_2 \b_3 \b_4] \a }) = 0 \,.
  \la{tttwo}
 \eea

Comparing (\ref{ttone}) and (\ref{tttwo}), we deduce that
$\nabla_{\a} F_{\b_1 \cdots \b_4} = 0$. From this it follows that
the $T^3$ component with five holomorphic indices vanishes, and this
implies that $(T^2_{\a \b})_{\bar
  \g \bar \d} = 0$. Therefore $T^2=0$.

   As we have already mentioned a direct inspection of  (\ref{holcon}) reveals that
  all $T^k$ are determined in terms of $T^1$ and $T^2$. Thus $T^k=0$
  and so ${\cal R}=0$.
Therefore, the reduced holonomy of $N=31$ backgrounds with an $SU(5)$-invariant normal
is $\{1\}$, and so these backgrounds are locally isometric to the maximally supersymmetric ones.

 \newsection{$(Spin(7)\ltimes\bR^8)\times\bR$-invariant normal}

\subsection{Integrability conditions}

The null case can be investigated in a similar way. For this we use the null basis of \cite{uggpdr} and choose the
normal spinor as
\bea
\nu=1+e_{1234}~.
\eea
A basis in the space of Majorana spinors orthogonal to $\nu$ is
\bea
&&1+e_{1234}, \quad i(1-e_{1234})~, \quad i(e_5-e_{12345})~,
\cr
&&e_\rho +{1 \over 3!} \epsilon^{\rho \sigma_1 \sigma_2 \sigma_3} e_{\sigma_1 \sigma_2 \sigma_3}~, \quad
i(e_\rho -{1 \over 3!} \epsilon^{\rho \sigma_1 \sigma_2 \sigma_3} e_{\sigma_1 \sigma_2 \sigma_3})~,
\cr
&&e_{\rho 5}+{1 \over 3!} \epsilon^{\rho \sigma_1 \sigma_2 \sigma_3} e_{\sigma_1 \sigma_2 \sigma_3 5}~ , \quad
i(e_{\rho 5} -{1 \over 3!} \epsilon^{\rho \sigma_1 \sigma_2 \sigma_3} e_{\sigma_1 \sigma_2 \sigma_3 5} )~,
\cr
&&i(e_{\rho_1 \rho_2} +{1 \over 2} \epsilon^{\rho_1 \rho_2 \mu_1 \mu_2} e_{\mu_1 \mu_2}), \quad
e_{\rho_1 \rho_2} -{1 \over 2} \epsilon^{\rho_1 \rho_2 \mu_1 \mu_2} e_{\mu_1 \mu_2}~,
\cr
&&i(e_{\rho_1 \rho_2 5} +{1 \over 2} \epsilon^{\rho_1 \rho_2 \mu_1 \mu_2} e_{\mu_1 \mu_2 5} ), \quad
e_{\rho_1 \rho_2 5}-{1 \over 2} \epsilon^{\rho_1 \rho_2 \mu_1 \mu_2} e_{\mu_1 \mu_2 5}~.
\eea
{}For the analysis we shall present below, it is convenient to introduce a new $SU(4)$-covariant basis as

\bea
&&\theta^\nat = i(e_5-e_{12345})~, \quad \theta^+ = i(1-e_{1234})~, \quad \theta^- = 1+e_{1234}~,
\cr
&&\theta^{-\rho} = {\sqrt{2} \over 3!} \epsilon^{\rho \sigma_1 \sigma_2 \sigma_3} e_{\sigma_1 \sigma_2 \sigma_3}~, \quad \theta^{-\bar{\rho}}
= \sqrt{2} e_\rho~,
\cr
 &&\theta^{\rho} = {\sqrt{2} \over 3!} \epsilon^{\rho \sigma_1 \sigma_2 \sigma_3} e_{\sigma_1 \sigma_2 \sigma_3 5}~ ,
\quad \theta^{\bar{\rho}} = \sqrt{2} e_{\rho 5}~,
\cr
&&\theta^{- \bar{\rho} \bar{\sigma}} = \sqrt{2}e_{\rho \sigma}~,
\quad \theta^{\bar{\rho} \bar{\sigma}} = \sqrt{2}e_{\rho \sigma 5}~,~~~\l,\m,\n,\r,\s=1,2,3,4~.
\la{su4basis}
\eea

It is then straightforward to show using (\ref{T-u-relation}) and the form bi-linears of appendix A that
\bea
&&u_\nat = 4i (T^2)_\mu{}^\mu~,~~~
u_- = -8 \sqrt{2} (T^1)_-~,~~~
u_+ = 2 \sqrt{2} i (T^3)_{-\mu}{}^\mu~,~~~
u_{-\rho} = 8 \sqrt{2} (T^2)_{-\rho}~,~~~
\cr
&&u_{\rho} = -16 (T^1)_{\rho}~,~~~
\epsilon_{\rho \sigma}{}^{\bar{\mu}_1 \bar{\mu}_2} u_{\bar{\mu}_1 \bar{\mu}_2} =8 \sqrt{2} (T^2)_{\rho \sigma}~,~~~
\epsilon_{\rho \sigma}{}^{\bar{\mu}_1 \bar{\mu}_2} u_{- \bar{\mu}_1 \bar{\mu}_2} =  8 (T^3)_{-\rho \sigma}~,
\eea
where the two-form indices of $u$ and $T^k$ have been suppressed.
In addition, we find that (\ref{T-u-relation}) implies the following relations between the $T^k$:
\bea
(T^1)_+= (T^1)_\nat &=&0~,
\cr
(T^2)_{+-} = (T^2)_{+\rho} = (T^2)_{+\nat}&=&0~,
\cr
(T^2)_{-\nat} &=& (T^1)_-~,
\cr
(T^2)_{\nat \rho} &=& -(T^1)_\rho~,
\cr
(T^2)_{\rho \bar{\sigma}} &=& {1 \over 4} (T^2)_\mu{}^\mu \delta_{\rho \bar{\sigma}}~,
\cr
(T^2)_{\rho \sigma} + {1 \over 2} \epsilon_{\rho \sigma}{}^{\bar{\mu}_1 \bar{\mu}_2} (T^2)_{\bar{\mu}_1 \bar{\mu}_2} &=&0~,
\cr
(T^3)_{+-\rho} &=& (T^1)_\rho~,
\cr
(T^3)_{+-\nat}=(T^3)_{+\nat \rho}=(T^3)_{+\rho \sigma} = (T^3)_{+\rho \bar{\sigma}} &=&0~,
\cr
(T^3)_{-\nat \rho} &=& -(T^2)_{- \rho}~,
\cr
(T^3)_{- \rho \bar{\sigma}} &=& {1 \over 4} (T^3)_{-\mu}{}^\mu \delta_{\rho \bar{\sigma}}~,
\cr
(T^3)_{\nat \rho \sigma} &=& (T^2)_{\rho \sigma}~,
\cr
(T^3)_{\nat \rho \bar{\sigma}} &=& {1 \over 4} (T^2)_\mu{}^\mu \delta_{\rho \bar{\sigma}}~,
\cr
(T^3)_{\sigma_1 \sigma_2 \sigma_3} &=& -2 \epsilon_{\sigma_1 \sigma_2 \sigma_3}{}^{\bar{\rho}} (T^1)_{\bar{\rho}}~,
\cr
(T^3)_{\sigma_1 \sigma_2 \bar{\rho}} &=& 2 \delta_{\bar{\rho} [\sigma_1} (T^1)_{\sigma_2]}~,
\cr
(T^3)_{-\rho \sigma} + {1 \over 2} \epsilon_{\rho \sigma}{}^{\bar{\mu}_1 \bar{\mu}_2} (T^3)_{- \bar{\mu}_1 \bar{\mu}_2} &=&0~,
\cr
(T^4)_{+-\nat \rho} &=& -(T^1)_{\rho}~,
\cr
(T^4)_{+-\rho \sigma} &=& (T^2)_{\rho \sigma}~,
\cr
(T^4)_{+-\rho \bar{\sigma}} &=& {1 \over 4} (T^2)_\mu{}^\mu \delta_{\rho \bar{\sigma}}~,
\cr
(T^4)_{+\nat \rho \sigma} = (T^4)_{+ \nat \rho \bar{\sigma}} = (T^4)_{+ \sigma_1 \sigma_2 \sigma_3} = (T^4)_{+ \sigma_1 \sigma_2 \bar{\rho}}&=&0~,
\cr
(T^4)_{- \nat \rho \sigma} &=& (T^3)_{-\rho \sigma}~,
\cr
(T^4)_{-\nat \rho \bar{\sigma}} &=& {1 \over 4} (T^3)_{-\mu}{}^\mu \delta_{\rho \bar{\sigma}}~,
\cr
(T^4)_{-\sigma_1 \sigma_2 \sigma_3} &=& 2 (T^2)_{- \bar{\rho}} \epsilon^{\bar{\rho}}{}_{\sigma_1 \sigma_2 \sigma_3}~,
\cr
(T^4)_{- \sigma_1 \sigma_2 \bar{\rho}} &=& 2 \delta_{\bar{\rho}[\sigma_1} (T^2)_{|-| \sigma_2]}~,
\cr
(T^4)_{\nat \sigma_1 \sigma_2 \sigma_3} &=& -2 (T^1)_{\bar{\rho}} \epsilon^{\bar{\rho}}{}_{\sigma_1 \sigma_2 \sigma_3}~,
\cr
(T^4)_{\nat \sigma_1 \sigma_2 \bar{\rho}} &=& -2 \delta_{\bar{\rho} [\sigma_1} (T^1)_{\sigma_2]}~,
\cr
(T^4)_{\sigma_1 \sigma_2 \sigma_3 \sigma_4} &=& {1 \over 2} (T^2)_\mu{}^\mu \epsilon_{\sigma_1 \sigma_2 \sigma_3 \sigma_4}~,
\cr
(T^4)_{\sigma_1 \sigma_2 \sigma_3 \bar{\rho}} &=& -3 \delta_{\bar{\rho} [\sigma_1} (T^2)_{\sigma_2 \sigma_3]}~,
\cr
(T^4)_{\rho_1 \rho_2 \bar{\sigma}_1 \bar{\sigma}_2} &=&0~,
\cr
(T^5)_{+- \nat \rho \sigma} &=& (T^2)_{\rho \sigma}~,
\cr
(T^5)_{+-\nat \rho \bar{\sigma}} &=& {1 \over 4} (T^2)_\mu{}^\mu \delta_{\rho \bar{\sigma}}~,
\cr
(T^5)_{+-\sigma_1 \sigma_2 \sigma_3} &=& 2 (T^1)_{ \bar{\rho}} \epsilon^{\bar{\rho}}{}_{\sigma_1 \sigma_2 \sigma_3}~,
\cr
(T^5)_{+-\sigma_1 \sigma_2 \bar{\rho}} &=& 2  \delta_{\bar{\rho} [\sigma_1} (T^1)_{ \sigma_2]}~,
\cr
(T^5)_{+\nat \sigma_1 \sigma_2 \sigma_3} = (T^5)_{+ \nat \sigma_1 \sigma_2 \bar{\rho}} &=&0~,
\cr
(T^5)_{+\sigma_1 \sigma_2 \sigma_3 \sigma_4}
= (T^5)_{+ \sigma_1 \sigma_2 \sigma_3 \bar{\rho}} = (T^5)_{+ \sigma_1 \sigma_2 \bar{\rho}_1 \bar{\rho}_2} &=&0~,
\cr
(T^5)_{- \nat \sigma_1 \sigma_2 \sigma_3} &=& -2  (T^2)_{-\bar{\rho}} \epsilon^{\bar{\rho}}{}_{\sigma_1 \sigma_2 \sigma_3}~,
\cr
(T^5)_{-\nat \sigma_1 \sigma_2 \bar{\rho}} &=& -2 \delta_{\bar{\rho} [\sigma_1} (T^2)_{|-|\sigma_2]}~,
\cr
(T^5)_{-\sigma_1 \sigma_2 \sigma_3 \sigma_4} &=& ({1 \over 2}(T^3)_{-\mu}{}^\mu +2 (T^1)_-) \epsilon_{\sigma_1 \sigma_2 \sigma_3 \sigma_4}~,
\cr
(T^5)_{-\sigma_1 \sigma_2 \sigma_3 \bar{\rho}} &=& -3 \delta_{\bar{\rho} [\sigma_1} (T^3)_{|-| \sigma_2 \sigma_3]}~,
\cr
(T^5)_{-\sigma_1 \sigma_2 \bar{\rho}_1 \bar{\rho}_2} &=& -2 \delta_{\sigma_1 [\bar{\rho}_1} \delta_{\bar{\rho}_2] \sigma_2} (T^1)_-~,
\cr
(T^5)_{\nat \sigma_1 \sigma_2 \sigma_3 \sigma_4} &=& {1 \over 2} (T^2)_\mu{}^\mu \epsilon_{\sigma_1 \sigma_2 \sigma_3 \sigma_4}~,
\cr
(T^5)_{\nat \sigma_1 \sigma_2 \sigma_3 \bar{\rho}} &=& -3 \delta_{\bar{\rho} [\sigma_1} (T^2)_{\sigma_2 \sigma_3]}~,
\cr
(T^5)_{\nat \sigma_1 \sigma_2 \bar{\rho}_1 \bar{\rho}_2} &=&0~,
\cr
(T^5)_{\sigma_1 \sigma_2 \sigma_3 \sigma_4 \bar{\rho}} &=& 2 \epsilon_{\sigma_1 \sigma_2 \sigma_3 \sigma_4} (T^1)_{ \bar{\rho}}~,
\cr
(T^5)_{\sigma_1 \sigma_2 \sigma_3 \bar{\rho}_1 \bar{\rho}_2} &=& -6 \delta_{\bar{\rho}_1 [\sigma_1}
\delta_{\sigma_2 |\bar{\rho}_2} (T^1)_{ \sigma_3]}~.
\label{Spin7-constraints}
\eea
Observe that all components $T^k$ of the supercurvature ${\cal R}$ are determined in terms of $T^1$, $T^2$ and $T^3$.

\subsection{Solving the conditions}

We shall now use (\ref{field-eqs}), (\ref{T-constraints}) and the explicit expressions of $T^k$ in terms of the physical fields
which can be found in \cite{jfgpa}
to show that ${\cal R}=0$. Since all the components of ${\cal R}$ in this case depend of $T^1$, $T^2$ and $T^3$, let us first show that
$T^1=0$.  Due to \eqref{Spin7-constraints} and the
skew-symmetry of $(T^1_{MN})_P$, the only possible non-vanishing components of $T^1$ up to complex conjugation are
$(T^1_{\rho_1 \rho_2})_{\rho_3}$, $(T^1_{\rho_1 \rho_2})_{\bar{\sigma}}$, $(T^1_{\rho \sigma})_-$,
$(T^1_{\rho \bar{\sigma}})_-$.

First consider the condition on $T^2$ in \eqref{T-constraints}. Taking $Q= \nat$, this implies that $T^1$ satisfies
 \begin{align}
  (T^1_{\rho_1 \rho_2})_{\sigma} =  (T^1_{\rho_1 \rho_2})_{\bar \sigma} = 0 \,, \quad
  (T^1_{\rho_1 \rho_2})_{-} = - \tfrac{1}{2} \epsilon_{\rho_1 \rho_2}{}^{\bar \sigma_1
  \bar \sigma_2} (T^1_{\bar \sigma_1 \bar \sigma_2})_{-} \,, \quad
  (T^1_{\rho \bar \sigma})_{-} = \tfrac{1}{4} (T^1_{\l}{}^\l)_{-} g_{\rho \bar \sigma} \,.
 \end{align}
Next turn to the conditions in
\eqref{field-eqs}. From  $(T^2_{\nat [-})_{\rho \bar{\sigma}]} -{1 \over 3} (T^4_{\nat N})_{- \rho
\bar{\sigma}}{}^N=0$, we find
\be
(T^1_{\rho \bar{\sigma}})_- = - (T^1_\mu{}^\mu)_- \delta_{\rho
  \bar{\sigma}} \,,
\ee
which implies that
\be
(T^1_{\rho \bar{\sigma}})_-=0 \,.
\ee
In addition $(T^1_{\rho_1 \rho_2})_- + {1 \over 2}
(T^3_{\rho_1 N})_{\rho_2 -}{}^N =0$ implies that
\be
(T^3_{{\rho}_1 \bar{\lambda}})^{\bar{\lambda}}{}_{{\rho}_2 -} = -2 (T^1_{\rho_1 \rho_2})_- +{1 \over 4}(T^3_{\rho_1 \rho_2})_{-\mu}{}^\mu~.
\ee
Combining this result with the condition $(T^2_{\rho [\sigma})_{- \nat]} -
{1 \over 3}(T^4_{\rho N})_{\sigma - \nat}{}^N =0$, which yields
\be
(T^3_{{\rho}_1 \bar{\lambda}})^{\bar{\lambda}}{}_{{\rho}_2 -} = - 4(T^1_{\rho_1 \rho_2})_- +{1 \over 4}(T^3_{\rho_1 \rho_2})_{-\mu}{}^\mu~,
\ee
we find $(T^1_{\rho \sigma})_-=0$. Hence $T^1=0$.

We now turn our attention to $T^2$. From \eqref{Spin7-constraints} and the
symmetry property in \eqref{T-constraints},  it follows that  $(T^2_{MN})_{P+}=(T^2_{P+})_{MN}=(T^2_{P\nat})_{MN}=0$.
Furthermore, $(T^2_{MN})_{\r\s}=(T^2_{\r\s})_{MN}$ are self-dual, and $(T^2_{MN})_{\r\bar\s}=(T^2_{\r\bar\s})_{MN}$
are determined in terms of the trace.

Let us first consider the case where all four indices are of $SU(4)$ type.
{}From $(T^2_{\rho N})_{\bar \sigma}{}^N = 0$ and $(T^3_{[\rho_1 \rho_2})_{\bar \sigma_1 \bar \sigma_2 \nat]} = 0$, we find
respectively
 \begin{align}
   (T^2_{\rho  \l})_{\bar \sigma}{}^\l = \tfrac{1}{16} g_{\rho \bar \sigma} (T^2
   \cont{\sigma}) \cont{\l} \,, \quad
   (T^2_{\rho_1 \rho_2})_{\bar \sigma_1 \bar \sigma_2} = - 2 (T^2_{\bar \sigma_1
  [\rho_1})_{a_2] \bar \sigma_2} \,.
 \end{align}
By taking the trace of the second equation, we conclude that these expressions
vanish. Hence the equations imply that
$(T^2_{\rho_1 \rho_2})_{\bar \sigma_1 \bar \sigma_2} = (T^2_{\rho \bar \sigma})_{\l \bar \d} =
0$.
Furthermore, $(T^3_{[\rho_1 \rho_2})_{\rho_3 \bar \sigma_1 \nat]} = 0$ implies that
$(T^2_{\rho \sigma})_{\l \bar \d} = 0$. Therefore $T^2$ with only $SU(4)$ indices vanishes.

Next we consider the case where one of the indices equals $-$.
{}From $(T^2_{\rho N})_{-}{}^N = 0$ and $(T^2_{-[\rho_1})_{\rho_2 \rho_3]} -
\tfrac{1}{3} (T^4_{-N})_{\rho_1 \rho_2 \rho_3}{}^N = 0$, we find that
 \begin{align}
  (T^2_{- \rho}) \cont{\sigma} = - 4 (T^2_{\rho  \l})_-{}^\l \,, \quad
  (T^2_{- [ \rho_1})_{\rho_2 \rho_3 ]} = \tfrac{1}{12} \epsilon_{\rho_1 \rho_2
  \rho_3}{}^{\bar \sigma} (T^2_{- \bar \sigma}) \cont{\l} \,.
 \label{T2-constraints}
 \end{align}
In addition, we explore the   relations of $T^3$ which arise from
$(T^3_{[ \nat  \rho_1})_{\rho_2 \rho_3  - ]} = 0$, $(T^3_{[ \nat \bar \sigma})_{\rho_1
  \rho_2  - ]} = 0$, $(T^2_{\rho [-})_{\sigma_1 \sigma_2 ]} -
\tfrac{1}{3} (T^4_{ \rho N})_{- \sigma_1 \sigma_2}{}^N = 0$ and $(T^2_{\rho [-})_{\bar \sigma_1 \bar \sigma_2 ]} -
\tfrac{1}{3} (T^4_{\rho N})_{-  \bar \sigma_1 \bar \sigma_2}{}^N = 0$ to find
 \begin{align}
  & (T^3_{\nat [ \rho_1})_{\rho_2 \rho_3 ] - } = 2 (T^2_{- [ \rho_1})_{\rho_2 \rho_3 ]} \,,
  \notag \\
  & (T^3_{\nat \bar \sigma})_{\rho_1 \rho_2 - } = 4 (T^2_{- [ \rho_1})_{\rho_2 ] \bar \sigma}
  + 2 (T^2_{- \bar \sigma})_{\rho_1 \rho_2} - 2 (T^3_{\nat [ \rho_1})_{\rho_2 ] \bar \sigma -
  } \,, \notag \\
  & (T^3_{\rho \nat})_{- \sigma_1 \sigma_2} = 2 (T^2_{\rho -})_{\sigma_1 \sigma_2} - 4 (T^2_{\rho
  [\sigma_1})_{|-| \sigma_2]} - \tfrac{1}{2} (T^2_{- \bar \l}) \cont{\d} \epsilon^{\bar
  \l}{}_{\rho \sigma_1 \sigma_2} \,, \notag \\
  & (T^3_{\rho \nat})_{- \bar \sigma_1 \bar \sigma_2} = 2 (T^2_{\rho -})_{\bar \sigma_1 \bar
  \sigma_2} - 4 (T^2_{\rho
  [\bar \sigma_1})_{|-| \bar \sigma_2]} - 2 (T^2_{\rho \l})_{- \d} \epsilon_{\bar \sigma_1
  \bar \sigma_2}{}^{\l \d} \,.
 \end{align}
{}From the two  expressions above with three holomorphic indices  it follows
that
 \begin{align}
  (T^2_{- [ \rho_1})_{\rho_2 \rho_3 ]} = \tfrac{1}{8} \epsilon_{\rho_1 \rho_2
  \rho_3}{}^{\bar \sigma} (T^2_{- \bar \sigma}) \cont{\l} \,.
 \end{align}
Combining this with \eqref{T2-constraints}, we conclude that these expressions
vanish, and therefore $(T^2_{-\rho})_{\sigma \bar \l} = 0$. Then, the definition for
$(T^3_{\nat \bar \sigma})_{\rho_1 \rho_2 - }$ and its complex conjugate
imply that $(T^2_{-\rho})_{\sigma \l} = 0$. Therefore  $T^2$ with three $SU(4)$ indices also vanishes.

The only remaining non-vanishing components are $(T^2_{- \rho})_{- \sigma}$ and $(T^2_{-
  \rho})_{- \bar \sigma}$. First, note that $(T^2_{-\rho})_{-\sigma}$ is symmetric in the interchange of $\rho$ and
$\sigma$, while in terms of $F$ it is given by
 \begin{align}
 (T^2_{-\rho})_{-\sigma} = (T^3_{-\rho})_{-\sigma\nat}=\tfrac{1}{6}\nabla_{-}F_{\rho -
 \sigma\nat} \,,
 \end{align}
which is skew-symmetric in the interchange and so $(T^2_{-\rho})_{-\sigma}=0$.  Similarly, $(T^2_{-\rho})_{-\bar \sigma} = (T^2_{- \bar \sigma})_{- \rho}$ while
 \begin{align}
 (T^2_{-\rho})_{-\bar\sigma} = (T^3_{-\rho})_{-\bar\sigma\nat}=\tfrac{1}{6}\nabla_{-}F_{\rho -
 \bar \sigma\nat} = - \tfrac{1}{6}\nabla_{-}F_{\bar \sigma -
 \rho\nat} = - (T^2_{- \bar \sigma})_{- \rho} \,.
 \end{align}
Hence this component also vanishes. Therefore we conclude that $T^2=0$.

It remains to consider $T^3$, and in particular the components $(T^3)_{\mu \nu -}$ and $(T^3)_{\mu \bar{\nu} -}$.
The vanishing of $(T^3_{MN})_{\mu \nu}{}^N$ for $M=-, \nat, \rho, \bar{\rho}$ implies that
\be
(T^3_{+-})_{\mu \nu -} = (T^3_{+\nat})_{\mu \nu -} = (T^3_{+ \rho})_{\mu \nu -} = (T^3_{+ \bar{\rho}})_{\mu \nu -}=0~.
\ee
{}From the vanishing of $(T^4_{MN})_{- \mu \nu}{}^N$ for $M=-, \rho, \bar{\rho}$, we also get
\be
(T^3_{- \nat})_{\mu \nu -} = (T^3_{\nat \rho})_{\mu \nu -} = (T^3_{\nat \bar{\rho}})_{\mu \nu -} =0~.
\ee

Next, note that
 \begin{align}
 \label{finalaux1}
  (T^3_{- \rho})_{- \bar \sigma_1 \bar \sigma_2} = (T^3_{-\bar \sigma_1})_{-\bar \sigma_2 \rho} = {1 \over 4} (T^3_{- \bar{\sigma}_1})_\rho{}^\rho{}_-
  g_{\rho \bar{\sigma_2}} \,,
 \end{align}
and on symmetrizing this expression in $\sigma_1, \sigma_2$ and taking the trace, we find $(T^3_{- \bar{\sigma}_1})_\mu{}^\mu{}_-=0$
and hence $(T^3_{- \rho})_{- \bar \sigma_1 \bar \sigma_2} = 0$.

Combining the Bianchi identity for $F$ with
 \begin{align}
  (T^3_{-\rho_1})_{\rho_2 \bar \sigma_1 \bar \sigma_2}=\tfrac{1}{6}(\nabla_- F_{\rho_1 \rho_2
  \bar \sigma_1 \bar \sigma_2}-\nabla_{\rho_1} F_{- \rho_2 \bar \sigma_1 \bar \sigma_2})=0 \,,
 \end{align}
we find that $\nabla_{\rho_1}F_{- \rho_2 \bar \sigma_1 \bar
  \sigma_2}=0$ and hence $(T^3_{\rho_1 \rho_2})_{-\bar \sigma_1 \bar \sigma_2}$ vanishes.
$(T^3_{\rho_1 \rho_2})_{\sigma_1 \sigma_2 -}=0$  due to the duality condition in (\ref{Spin7-constraints}).
Finally, the Bianchi
identity for $F$ together with
 \begin{align}
 & (T^3_{-\rho})_{\l_1 \l_2 \bar \sigma}=\tfrac{1}{6}(\nabla_{-}F_{\rho \l_1\l_2 \bar
  \sigma}-\nabla_{\rho}F_{- \l_1 \l_2 \bar \sigma})=0 \,, \notag \\
 & (T^3_{\bar \sigma -})_{\rho \l_1 \l_2}=\tfrac{1}{6}(\nabla_{\bar \sigma}F_{-\rho \l_1
  \l_2}-\nabla_- F_{\bar \sigma \rho \l_1 \l_2 })=0 \,,
 \end{align}
imply that $\nabla_{\rho} F_{- \mu \nu \bar{\sigma}} = \nabla_{\bar{\sigma}} F_{- \rho \mu \nu} =0$, and
hence $(T^3_{\rho\bar \sigma})_{-\l_1 \l_2}=0$. Hence $(T^3)_{\mu \nu -}=0$.

In order to show that the remaining components of $(T^3)_{\mu \bar{\nu} -}$ also vanish,
note that  $(T^3_{MN})_{\mu \bar{\nu}}{}^N=0$ for $M=-, \nat, \rho, \bar{\rho}$ in (\ref{field-eqs}) implies
\be
(T^3_{+-})_{\mu \bar{\nu} -} = (T^3_{+\nat})_{\mu \bar{\nu} -} = (T^3_{+ \rho})_{\mu \bar{\nu} -}
= (T^3_{+ \bar{\rho}})_{\mu \bar{\nu} -}=0~,
\ee
and the vanishing of $(T^4_{MN})_{- \mu \bar{\nu}}{}^N$ for $M=-, \rho$ implies
\be
(T^3_{- \nat})_{\mu \bar{\nu} -} = (T^3_{\nat \rho})_{\mu \bar{\nu} -} =0 \ .
\ee

Next, as we have shown that $\nabla_{\rho} F_{- \mu \nu \bar{\sigma}} =0$, this implies $(T^3_{\rho_1 \rho_2})_{\mu \bar{\nu} -}=0$,
and hence $(T^3_{\bar{\rho}_1 \bar{\rho}_2})_{\mu \bar{\nu} -} =0$.
Also, $(T^3_{-\bar \sigma_1})_{-\bar \sigma_2 \rho} = 0$ from ({\ref{finalaux1}}).
Lastly, by taking traces
of the constraint $(T^3_{[\rho_1 \bar{\rho}_2})_{\sigma_1 \bar{\sigma}_2 -]}=0$ and using
$(T^3_{\rho_1 \sigma_1})_{\bar{\rho}_2 \bar{\sigma}_2 -}= (T^3_{\bar{\rho}_2 \bar{\sigma}_2})_{\rho_1 \sigma_1 -}=0$, we find
$(T^3_{\rho_1 \bar{\rho}_2})_{\sigma_1 \bar{\sigma}_2 -}=0$.
 Hence $(T^3)_{\mu \bar{\nu} -}=0$. These conditions are
then sufficient to show that $T^3=0$.

As we have already mentioned,  $T^k$ are determined from $T^1, T^2$ and $T^3$. Since $T^1=T^2=T^3=0$, $T^k=0$ and so ${\cal R}=0$.
Therefore, the reduced holonomy of $N=31$ backgrounds with a $(Spin(7)\ltimes\bR^8)\times\bR$-invariant normal is $\{1\}$, and so these
backgrounds are locally isometric to maximally supersymmetric ones. Combining this result with that of the previous section
section, we conclude that all $N=31$ backgrounds of eleven-dimensional supergravity admit locally an additional Killing spinor
and so they are maximally supersymmetric.

\newsection{$N=15$ in type I supergravities}

The non-existence of $N=15$ supersymmetric backgrounds in type I supergravities can be easily seen by combining the results
of \cite{iibpreons} and \cite{phgpug}. In particular, the normal to the 15 Killing spinors has stability subgroup
$Spin(7)\ltimes\bR^8$.  So there is only one case to consider. It is convenient to choose
\bea
\nu=e_2-e_{134}~.
\eea
Then combining the conditions of the backgrounds with Killing spinors that have stability subgroup $\bR^8$ and
those that have stability subgroup $G_2$ in \cite{phgpug}, one finds that the dilaton $\Phi$ is constant and
the non-vanishing components of $H$ are $H_{-ij}$, where $i,j=1,\dots 8$. The dilatino Killing spinor equation
becomes
\bea
H_{-ij} \Gamma^{-ij}\e^r=0~.
\eea
The existence of a non-trivial solution for this equation is equivalent
 to requiring that there are seven linearly independent spinors in the chiral or anti-chiral representation  of $Spin(8)$, depending on conventions,
 with a non-trivial stability subgroup. This is not the case and so $H=0$.
 Similarly,  the integrability condition of the gravitino Killing spinor equation implies that the supercovariant curvature
 of the connection with torsion vanishes, $\hat R=0$. Since $H=0$, $\hat R=R=0$, the Riemann curvature of the spacetime
 vanishes. The rest of the fluxes, e.g.~gauge field strengths, can also be shown to vanish.
 Therefore, the spacetime is locally isometric to Minkowski space with constant dilaton,
 and vanishing three-form and gauge field fluxes.

\newsection{Concluding remarks}

We have shown that eleven-dimensional supergravity backgrounds with 31 supersymmetries are locally
isometric to maximally supersymmetric ones. This result together with that of \cite{jfgpa} (locally) classify the supersymmetric backgrounds
of eleven-dimensional supergravity with $N=31$ and $N=32$ supersymmetries.  The Killing spinor equations
 of eleven-dimensional
supergravity  for the $N=1$  backgrounds have been solved in  \cite{gjs}. So far, these are the only three
cases in eleven-dimensions that the geometry of the  backgrounds has been identified for a given $N$.
Furthermore,  the result of this paper together with those obtained in
\cite{iibpreons} and  \cite{bandos} rule out the existence of $N=31$   solutions in  eleven- and type II ten-dimensional  supergravities.
In addition, a straightforward
argument
can rule out the existence of $N=15$
backgrounds in type I ten-dimensional supergravities. In lower-dimensions, a similar conclusion has been reached
for the cases that have been investigated in \cite{jgutowski}. There are many more lower dimensional cases
that can be explored.

It is clear from the cases that have been examined so far that backgrounds with $N_{\rm max}-1$ number of supersymmetries are severely
restricted. This   raises the possibility that there are much less
supersymmetric backgrounds in ten and eleven dimensions  than those that may have been expected from
the holonomy argument of \cite{hull, duff, gpdta, gpdtb}. In the proof that the $N=31$ eleven-dimensional backgrounds admit $32$ supersymmetries, we have
used both the conditions that arise from the Killing spinor equations as well as field equations and Bianchi identities.
It has been the field equations and Bianchi identities that enforced the condition that the supercovariant curvature vanishes -- the conditions
arising from the Killing spinor equations were not sufficient. Dynamical information has been necessary to construct the proof.
This is unlike the type II theories where the Killing spinor equations were sufficient to establish the result.

Another property of the $N=31$ backgrounds in eleven or ten dimensions is that
the stability subgroup of Killing
spinors in $Spin(10,1)$ or $Spin(9,1)$ is trivial, i.e.~${\rm
  stab}(\e)=\{1\}$. These are the first examples, other than those with maximal supersymmetry, that have this
  property. It is encouraging that it turned out to be that such
backgrounds are in fact maximally supersymmetric. This may suggest that even
backgrounds with a small number of Killing spinors but with a trivial stability subgroup in the gauge
group of the Killing spinor equations are severely restricted, though it is
possible that  such new backgrounds exist. If this is the case, the classification of supersymmetric backgrounds in ten and
 eleven dimensions may be somewhat  simplified. It would be worth investigating more such examples in the future.

\vskip 0.5cm
\noindent{\bf Acknowledgements} \vskip 0.1cm Part of this work was completed while D.R.~was a post-doc at King's College
London, for which he would like to acknowledge the PPARC grant
PPA/G/O/2002/00475. In addition, he is presently supported by the European
EC-RTN project MRTN-CT-2004-005104, MCYT FPA 2004-04582-C02-01 and CIRIT GC
2005SGR-00564. U.G.~has a
postdoctoral fellowship funded by the Research Foundation
K.U.~Leuven.

\vskip 0.5cm

\setcounter{section}{0}
\appendix {Spacetime form spinor bi-linears}

In the computation of the conditions that arise from the integrability condition
${\cal R}\e^r=0$, we have used the form spinor bi-linears of the $SU(5)$-invariant  normal spinor $\nu$
and a basis $\theta^r$, (\ref{SU(5)-basis}),  that spans the 31 Killing spinors. These bi-linears are defined as
\bea
\tau^r={1\over k!} B(\theta^r,
\Gamma_{A_1A_2\dots A_k}\nu)\, e^{A_1}\wedge e^{A_2}\wedge \dots\wedge e^{A_k}~.
\eea
 In particular the non-vanishing components of the one-forms are
\bea
\t_\a^\b=2\d_\a^\b~,~~~\t_0^0=-2~.
\eea
The two-forms are
\bea
\t_{0\a}^\b=-2i\d_\a^\b~,~~~\t_{\a\bar\b}^0=-2 i
g_{\a\bar\b}~,~~~\t_{\a\b}^{\g\d}=4\sqrt{2} i \d_{[\a\b]}^{\g\d}~.
\eea
The three-forms are
\bea
\t_{0\a\b}^{\g\d}=-4\sqrt{2}\d_{[\a\b]}^{\g\d}~,~~~\t_{\g_1\g_2\g_3}^{\bar\a\bar\b}=4i\e_{\g_1\g_2\g_3}{}^{\bar\a\bar\b}~,~~~\t_{\a\b\bar\g}^{\d}=-4\d_{[\a}^{\d}g_{\b]\bar\g}~.
\eea
The four-forms are
\bea
&&\t_{0\g_1\g_2\g_3}^{\bar\a\bar\b}=4\e_{\g_1\g_2\g_3}{}^{\bar\a\bar\b}~,~~~\t_{0\b_1\b_2\bar\g}^{\a}=4i
\d_{[\b_1}^{\a}g_{\b_2]\bar\g}~,~~~\t_{\b_1\b_2\b_3\b_4}^{\bar\a}=4\sqrt{2}\e_{\b_1\b_2\b_3\b_4}{}^{\bar\a}~,
\cr
&&~~~~~~~~~~~~~~~~~~~~~\t_{\g_1\g_2\g_3\bar\g_4}^{\a\b}=-12\sqrt{2}i
\d_{[\g_1\g_2}^{\a\b}g_{\g_3]\bar\g_4}~,
\eea
and the five-forms are
\bea
&&\t_{0\b_1\b_2\b_3\b_4}^{\bar\a}=4\sqrt{2}i
\e_{b_1\b_2\b_3\b_4}{}^{\bar\a}~,~~~\t_{0\g_1\g_2\g_3\bar\g_4}^{\a\b}=12\sqrt{2}\d_{[\g_1\g_2}^{\a\b}g_{\g_3]\bar\g_4}~,~~~
\t_{0\a\b\bar\g\bar\d}^0=4g_{\a[\bar\g}g_{|\b|\bar\d]}~,
\cr
&&\t_{\a_1\a_2\a_3\a_4\a_5}^0=-4\sqrt{2}i \e_{\a_1\a_2\a_3\a_4\a_5}~,~~~
\t_{\b_1\b_2\b_3\b_4\bar\g}^{\bar\a_1\bar\a_2}=8i
\e_{\b_1\b_2\b_3\b_4}{}^{[\bar\a_1}\d_{\bar\g}^{\bar\a_2]}~,~~~
\cr
&&\t_{\b_1\b_2\b_3\bar\g_1\bar\g_2}^{\a}=-12
\d_{[\b_1}^{\a}g_{\b_2|\bar\g_1|}g_{\b_3]\bar\g_2}~,
\eea
where we have used $\delta_{[\a_1 \a_2]}^{\b_1 \b_2} =  \delta_{[\a_1}^{\b_1} \delta_{\a_2]}^{\b_2}$.

Similarly, in the computation of the integrability conditions
of $N=31$ backgrounds with a $(Spin(7)\ltimes \bR^8)\times \bR$-invariant normal spinor $\nu$, we have used the spacetime form spinor bi-linears
of $\nu$ with the elements of the spinor basis (\ref{su4basis}). In particular, we find that
the one-forms are
\be
\tau^-{}_- = -2 \sqrt{2}, \quad \tau^{\rho}{}_\sigma = -2 \delta^\rho_\sigma~,
\ee
the two-forms are
\bea
&&\tau^\nat{}_{\rho \bar{\sigma}} = 2i \delta_{\rho \bar{\sigma}}, \quad \tau^-{}_{-\nat} = 2 \sqrt{2},
\quad \tau^{-\rho}{}_{- \sigma} = -2 \sqrt{2} \delta^\rho_\sigma~,
\cr
&&\tau^{\rho}{}_{\nat \sigma} = -2 \delta^\rho_\sigma, \quad \tau^{\bar{\rho} \bar{\sigma}}{}_{\bar{\mu}_1 \bar{\mu}_2} =4 \sqrt{2} \delta^{\bar{\rho}
\bar{\sigma}}_{\bar{\mu}_1 \bar{\mu}_2},
\quad \tau^{\bar{\rho} \bar{\sigma}}{}_{{\mu}_1 {\mu_2}} = -2 \sqrt{2} \epsilon^{\bar{\rho} \bar{\sigma}}{}_{{\mu}_1 {\mu}_2}~,
\eea
the three-forms are
\bea
&&\tau^\nat{}_{\nat \rho \bar{\sigma}} = -2i \delta_{\rho \bar{\sigma}}, \quad \tau^+{}_{- \rho \bar{\sigma}} = -2 \sqrt{2} i \delta_{\rho \bar{\sigma}},
\quad \tau^{-\rho}{}_{- \nat \sigma} = -2 \sqrt{2} \delta^\rho_\sigma~,
\cr
&&\tau^{\rho}{}_{+-\sigma} =-2 \delta^\rho_\sigma, \quad \tau^{\rho}{}_{\sigma_1 \sigma_2 \bar{\lambda}}
=-4 \delta_{\bar{\lambda} [\sigma_1} \delta^\rho_{\sigma_2]}, \quad \tau^{\rho}{}_{\bar{\sigma}_1
\bar{\sigma}_2 \bar{\sigma}_3} = -4 \epsilon^\rho{}_{\bar{\sigma}_1
\bar{\sigma}_2 \bar{\sigma}_3}~,
\cr
&&\tau^{- \bar{\rho} \bar{\sigma}}{}_{- \bar{\mu}_1 \bar{\mu}_2} = -8 \delta^{\bar{\rho} \bar{\sigma}}_{\bar{\mu}_1 \bar{\mu}_2},
\quad \tau^{- \bar{\rho} \bar{\sigma}}{}_{- {\mu}_1
{\mu}_2} = 4 \epsilon^{\bar{\rho} \bar{\sigma}}{}_{{\mu}_1 {\mu}_2}~,
\cr
&&\tau^{\bar{\rho} \bar{\sigma}}{}_{\nat \bar{\mu}_1 \bar{\mu}_2} = -4 \sqrt{2} \delta^{\bar{\rho} \bar{\sigma}}_{\bar{\mu}_1 \bar{\mu}_2},
\quad \tau^{\bar{\rho} \bar{\sigma}}{}_{\nat {\mu}_1 {\mu}_2} = 2 \sqrt{2} \epsilon^{\bar{\rho} \bar{\sigma}}{}_{{\mu}_1 {\mu}_2}~,
\eea
the four-forms are
\bea
&&\tau^\nat{}_{+-\rho \bar{\sigma}} =2i \delta_{\rho \bar{\sigma}}, \quad \tau^\nat{}_{\rho_1 \rho_2 \rho_3 \rho_4}=
4i \epsilon_{\rho_1 \rho_2 \rho_3 \rho_4}, \quad \tau^+{}_{- \nat \rho \bar{\sigma}}= 2 \sqrt{2} i \delta_{\rho \bar{\sigma}}~,
\cr
&&\tau^{-\rho}{}_{- \sigma_1 \sigma_2 \bar{\lambda}} =-4 \sqrt{2} \delta_{\bar{\lambda} [\sigma_1} \delta^\rho_{\sigma_2]},
\quad \tau^{-\rho}{}_{- \bar{\sigma}_1 \bar{\sigma}_2 \bar{\sigma}_3} = -4 \sqrt{2} \epsilon^\rho{}_{\bar{\sigma}_1 \bar{\sigma}_2 \bar{\sigma}_3}~,
\cr
&&\tau^{\rho}{}_{+-\nat \sigma}=-2 \delta^\rho_\sigma, \quad \tau^{\rho}{}_{\nat \sigma_1 \sigma_2 \bar{\lambda}}
= -4 \delta_{\bar{\lambda} [\sigma_1} \delta^\rho_{\sigma_2]}, \quad \tau^{\rho}{}_{\nat  \bar{\sigma}_1 \bar{\sigma}_2 \bar{\sigma}_3}
= -4  \epsilon^\rho{}_{\bar{\sigma}_1 \bar{\sigma}_2 \bar{\sigma}_3}~,
\cr
&&\tau^{- \bar{\rho} \bar{\sigma}}{}_{-\nat \bar{\mu}_1 \bar{\mu}_2} = 8 \delta^{\bar{\rho} \bar{\sigma}}_{\bar{\mu}_1 \bar{\mu}_2}, \quad
\tau^{- \bar{\rho} \bar{\sigma}}{}_{- \nat {\mu}_1 {\mu}_2} = -4\epsilon^{\bar{\rho} \bar{\sigma}}{}_{{\mu}_1 {\mu}_2}~,
\cr
&&\tau^{\bar{\rho} \bar{\sigma}}{}_{+- \bar{\mu}_1 \bar{\mu}_2} =4 \sqrt{2} \delta^{\bar{\rho} \bar{\sigma}}_{\bar{\mu}_1 \bar{\mu}_2},
\quad \tau^{\bar{\rho} \bar{\sigma}}{}_{+-{\mu}_1 {\mu}_2} =-2 \sqrt{2}\epsilon^{\bar{\rho} \bar{\sigma}}{}_{{\mu}_1 {\mu}_2}~,
\cr
&&\tau^{\bar{\rho} \bar{\sigma}}{}_{\bar{\sigma}_1 \bar{\sigma}_2 \bar{\sigma}_3 {\lambda}} = -12 \sqrt{2} \delta_{{\lambda} [ \bar{\sigma}_1}
\delta^{\bar{\rho}}_{\bar{\sigma}_2} \delta^{\bar{\sigma}}_{\bar{\sigma}_3]},
\quad \tau^{\bar{\rho} \bar{\sigma}}{}_{\bar{\lambda} {\sigma}_1 {\sigma}_2 {\sigma}_3} = 4 \sqrt{2} \epsilon_{{\sigma}_1 {\sigma}_2 {\sigma}_3}
{}^{[\bar{\rho}} \delta^{\bar{\sigma}]}_{\bar{\lambda}}~,
\eea
and the five-forms are
\bea
&&\tau^\nat{}_{+-\nat \rho \bar{\sigma}} =-2i \delta_{\rho \bar{\sigma}}, \quad \tau^\nat{}_{\nat \sigma_1 \sigma_2 \sigma_3 \sigma_4}
= -4i \epsilon_{\sigma_1 \sigma_2 \sigma_3 \sigma_4}~,
\cr
 &&\tau^-{}_{- \sigma_1 \sigma_2 \bar{\rho}_1 \bar{\rho}_2}
=4 \sqrt{2} \delta_{\sigma_1 [\bar{\rho}_1} \delta_{\bar{\rho}_2] \sigma_2}, \quad \tau^+{}_{- {\sigma}_1
{\sigma}_2 {\sigma}_3 {\sigma}_4} = -4 \sqrt{2}i  \epsilon_{{\sigma}_1
{\sigma}_2 {\sigma}_3 {\sigma}_4}~,
\cr
&&\tau^-{}_{- {\sigma}_1
{\sigma}_2 {\sigma}_3 {\sigma}_4} = -4 \sqrt{2}  \epsilon_{{\sigma}_1
{\sigma}_2 {\sigma}_3 {\sigma}_4}~,
\cr
&&\tau^{-\rho}{}_{- \nat \sigma_1 \sigma_2 \bar{\lambda}} = -4 \sqrt{2}\delta_{\bar{\lambda} [\sigma_1} \delta^\rho_{\sigma_2]}, \quad
\tau^{-\rho}{}_{- \nat \bar{\sigma}_1 \bar{\sigma}_2 \bar{\sigma}_3} = -4 \sqrt{2} \epsilon^\rho{}_{\bar{\sigma}_1 \bar{\sigma}_2 \bar{\sigma}_3}~,
\cr
&&\tau^{\rho}{}_{+- \sigma_1 \sigma_2 \bar{\lambda}} = -4 \delta_{\bar{\lambda} [\sigma_1} \delta^\rho_{\sigma_2]}, \quad
\tau^{\rho}{}_{+- \bar{\sigma}_1 \bar{\sigma}_2 \bar{\sigma}_3} = -4 \epsilon^\rho{}_{\bar{\sigma}_1 \bar{\sigma}_2 \bar{\sigma}_3}~,
\cr
&&\tau^{\rho}{}_{\sigma_1 \sigma_2 \sigma_3 \bar{\rho}_1 \bar{\rho}_2} = 12 \delta_{\bar{\rho}_1 [\sigma_1}
\delta_{\sigma_2 |\bar{\rho}_2|} \delta^\rho_{\sigma_3]}, \quad
\tau^{\rho}{}_{\lambda \bar{\sigma}_1 \bar{\sigma}_2 \bar{\sigma}_3 \bar{\sigma}_4} = -4 \delta^\rho_\lambda
\epsilon_{ \bar{\sigma}_1 \bar{\sigma}_2 \bar{\sigma}_3 \bar{\sigma}_4}~,
\cr
&&\tau^{-\bar{\rho} \bar{\sigma}}{}_{- \bar{\sigma}_1 \bar{\sigma}_2 \bar{\sigma}_3 {\lambda}} =  24
\delta_{{\lambda} [\bar{\sigma}_1} \delta^{\bar{\rho}}_{\bar{\sigma}_2}
\delta^{\bar{\sigma}}_{\bar{\sigma}_3]}, \quad \tau^{- \bar{\rho} \bar{\sigma}}{}_{- \bar{\lambda} {\sigma}_1 {\sigma_2} {\sigma}_3}
=-8 \epsilon_{{\sigma}_1 {\sigma_2} {\sigma}_3}{}^{[\bar{\rho}} \delta^{\bar{\sigma}]}_{\bar{\lambda}}~,
\cr
&&\tau^{\bar{\rho} \bar{\sigma}}{}_{+-\nat \bar{\mu}_1 \bar{\mu}_2} = -4 \sqrt{2} \delta^{\bar{\rho} \bar{\sigma}}_{\bar{\mu}_1 \bar{\mu}_2}, \quad
\tau^{\bar{\rho} \bar{\sigma}}{}_{+-\nat {\mu}_1 {\mu}_2} = 2 \sqrt{2} \epsilon^{\bar{\rho} \bar{\sigma}}{}_{{\mu}_1 {\mu}_2}~,
\cr
&&\tau^{\bar{\rho} \bar{\sigma}}{}_{\nat \bar{\sigma}_1 \bar{\sigma}_2 \bar{\sigma}_3 {\lambda}} = 12 \sqrt{2}
\delta_{{\lambda} [\bar{\sigma}_1} \delta^{\bar{\rho}}_{\bar{\sigma}_2}
\delta^{\bar{\sigma}}_{\bar{\sigma}_3]}, \quad
\tau^{\bar{\rho} \bar{\sigma}}{}_{\nat \bar{\lambda} {\sigma}_1 {\sigma_2} {\sigma}_3}
=-4 \sqrt{2}\epsilon_{{\sigma}_1 {\sigma_2} {\sigma}_3}{}^{[\bar{\rho}} \delta^{\bar{\sigma}]}_{\bar{\lambda}}~.
\eea
The components of $\tau^{\bar{\rho}}$ and $\tau^{- \bar{\rho}}$ are obtained from the above expressions
by complex conjugation.

\end{document}